\documentclass[conference]{IEEEtran}
\IEEEoverridecommandlockouts
\usepackage{cite}
\usepackage{amsmath,amssymb,amsfonts}
\usepackage{algorithmic}
\usepackage{graphicx}
\usepackage{textcomp}
\usepackage{xcolor}
\usepackage{tcolorbox}
\usepackage{subcaption}
\usepackage{hyperref}
\usepackage{multirow}
\usepackage{enumerate}
\usepackage{enumitem}
\usepackage{subcaption}
\usepackage{todonotes}
\usepackage{array}
\usepackage[T1]{fontenc}
\usepackage{times}

\def\BibTeX{{\rm B\kern-.05em{\sc i\kern-.025em b}\kern-.08em
    T\kern-.1667em\lower.7ex\hbox{E}\kern-.125emX}}

\newcommand{\dsname}[0]{{\small\tt CoDocBench}}
\newcommand{\dsnametitle}[0]{{\tt CoDocBench}}

\begin{document}
\title{\dsnametitle: A Dataset for Code-Documentation Alignment in Software Maintenance}

%


\author{
\IEEEauthorblockN{Kunal Pai, Premkumar Devanbu, and Toufique Ahmed \\
University of California, Davis, USA
}}


\maketitle

\begin{abstract}
One of the central tasks in software maintenance is being able to understand and develop code changes.  
Thus, given a natural language description of the desired new operation of a function,
an agent (human or AI) might be asked to
generate the set of edits to that function to implement the desired new operation; 
likewise, given a set of edits to a function,
an agent might be asked to generate a changed description, of that function's new workings. 
Thus, there is 
an incentive to train a neural model for change-related tasks. Motivated by this, we offer
a new, ``natural", large dataset of \emph{coupled changes to code and documentation} mined from actual
high-quality GitHub projects, where each sample represents a single commit where the code \emph{and}
the associated docstring were changed \emph{together}. 
We present the methodology for gathering
the dataset, and some sample, challenging (but realistic) tasks where our dataset provides
opportunities for both learning and evaluation. We find that
current models (specifically Llama-3.1 405B, Mixtral 8$\times$22B) do
find these maintenance-related tasks challenging. 
\end{abstract}


%
\IEEEpeerreviewmaketitle


\section{Introduction}
Software maintenance activities are reported to consume 60-80\% of
overall software budgets~\cite{canfora2001software}, and thus
constitute an attractive target for efforts
to manage costs. As language
models applications to software engineering
tasks continue to burgeon, the value of LLM for software maintenance has
been recognized. Various applications of LLM
to maintenance have been promoted, such as
automated code repair~\cite{xia2023automated}, 
automated response to code review comments~\cite{frommgen2024resolving}, 
industrial-scale generalized code maintenance support~\cite{didact}, and
even automatic response to submitted GitHub issues~\cite{zhang2024autocoderover}, with proposed changes.  

Based as they are on machine-learning, all the above depend on realistic, well-curated datasets. 
Our goal in this work is to propose a new dataset,
for a specific aspect of software maintenance $\ldots$ \emph{supporting well-documented code changes}. It has
been reported that documentation practice in industry really needs improvement~\cite{visconti2002overview}, and that
many maintenance difficulties arise as a result. 
Our work is focused on the specific problem in code documentation raised in Schrek \emph{et al}.~\cite{schreck2007documentation}, \emph{viz.,}
how code and the associated natural language
description (DocStrings) are not always 
kept up to date; often code is changed, but the documentation isn't. Schreck \emph{et al}. note that Docstrings are updated  only around
33\% of the time that code is changed; this phenomenon suggests that LLMs could help programmers better document changes. 

\begin{figure}
    \centering
    \includegraphics[width=\linewidth]{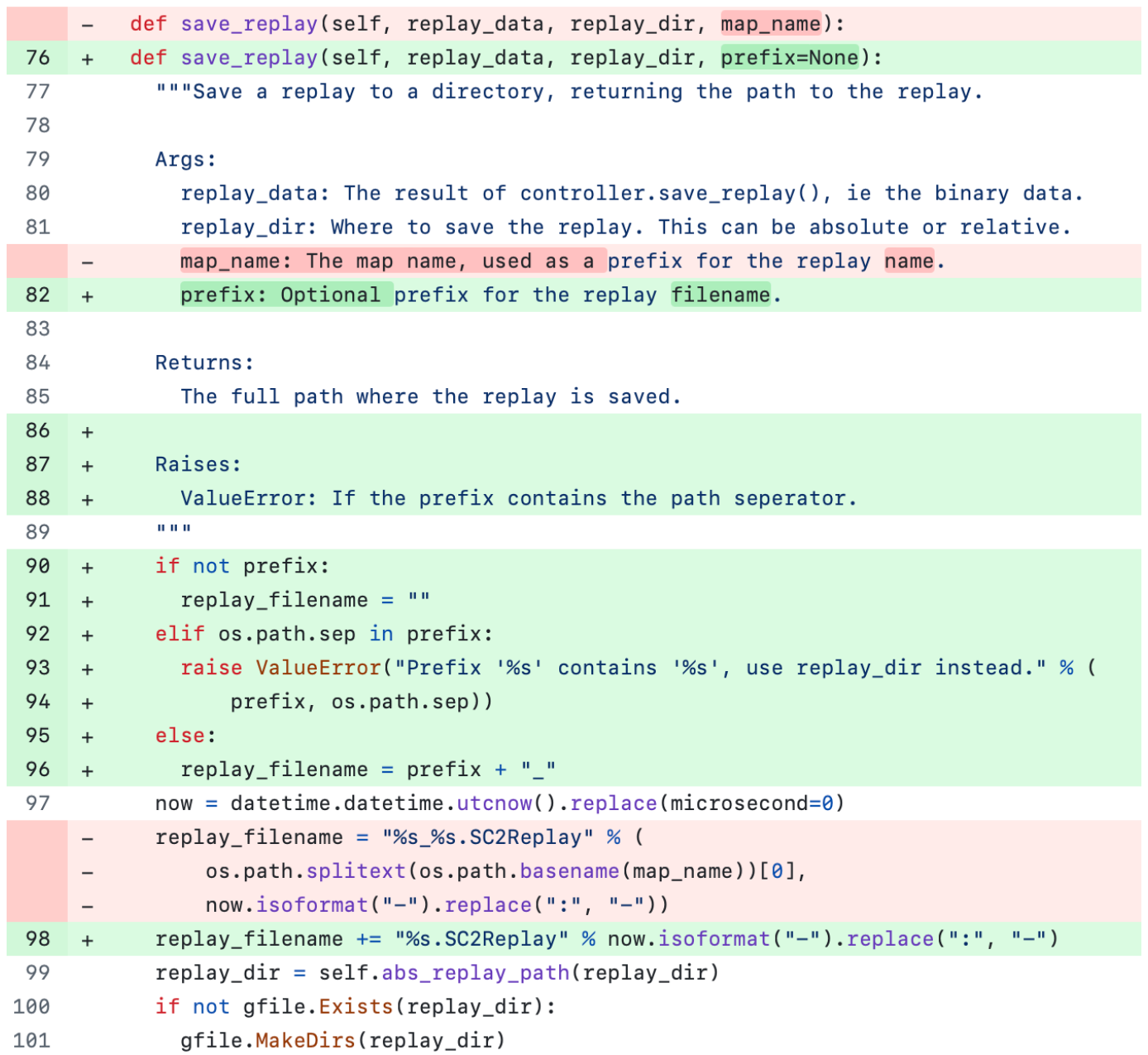}
    \caption{An example where docstring and function were changed simultaneously in the same commit}
    \label{fig:example}
\end{figure}

We introduce a new dataset, \dsname, aimed at training and evaluating language models in tasks related to helping developers better couple code and document changes. ~\autoref{fig:example} presents an example where an argument of the function is dropped and docstring-function pair were updated to accommodate the change.  
Specifically, we envision two initial tasks based
on this dataset. First, given old code, old associated docstring, and new version of the code: create a new docstring better aligned with the new code; second, given old docstring, old associated code, and a new docstring (reflecting the intended new function of the code), generate new implementation aligned with the new docstring. If such tasks could be automated, one can expect that the poorly-coupled Code-Documentation update rate of 33\% reported by Schrek \emph{et al}. ~\cite{schreck2007documentation} could be improved, thus leading to better documented and more maintainable code.

\dsname{} consists of $4573$ high-quality samples of coupled code-documentation
changes from GitHub; we have selected changes
where developers have indeed changed both the code
and the documentation \emph{in the same commit}.  In
the following we describe: 1) the collection \& curation procedure
2) the dataset \emph{per se}, and 3) and some illustrative studies
of the data.  
\begin{figure*}[h]
    \centering
    \includegraphics[width=.9\textwidth]{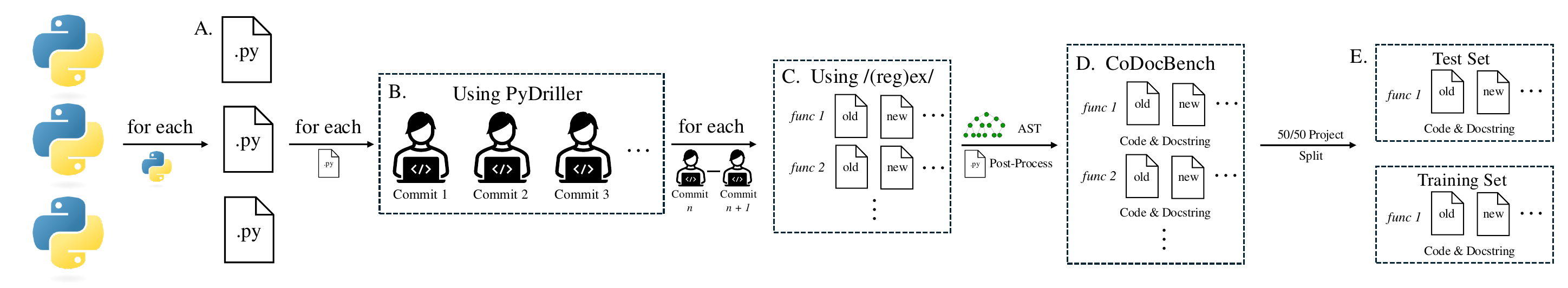}
    \caption{Workflow for creating \dsnametitle. The figure illustrates a multi-stage process: (A) Collect all  Python files per project, (B) Commit processing per Python file, (C) Using regular expressions for pairwise commit change detection per function, (D) Using AST and post-processing to refine detected changes and create \dsnametitle, and (E) Generating a test and training set through a random 50/50 project split.}
    \label{fig:dataset-collection-methodology}
\end{figure*}

\section{Dataset Collection Methodology}

~\autoref{fig:dataset-collection-methodology} presents the overall flow process of dataset construction. We begin with a curated list of the top 200 Python projects on GitHub to ensure a diverse and representative sample. These projects are selected based on criteria such as having a high star count (to reflect popularity and community interest) and recent commit activity (to ensure the projects are actively maintained). 
For each selected project, we gather all associated Python source files. Using PyDriller~\cite{spadini2018pydriller}, a Python framework for mining Git, we iterated over all commits that impact each Python file to trace the development history.
PyDriller can efficiently 
process Git data, allowing rapid analysis across a large number of commits over numerous files. We identified instances where the same function was modified in consecutive commits and recorded both the docstrings and code snippets from these commits. The detection of changes was implemented using the Python \texttt{re} package~\cite{friedl2006mastering}, which enabled us to effectively identify modifications. We specifically used regular expressions to efficiently detect function definitions (lines starting with ``\texttt{def}'') and text enclosed within triple quotes inside functions, allowing us to separate docstrings from the actual code.
As a post-processing step, we excluded data points where only the docstring or only the code had been altered, to ensure that 
our dataset contained only coupled-change instances where \emph{both} code and docstrings were modified together.
Entries involving only whitespace changes in either code or docstrings were also excluded.
Finally, we validated the extracted data using Python's Tree-Sitter~\cite{tree-sitter} and the \texttt{function\_parser} package provided as part of the CodeSearchNet Challenge~\cite{husain2019codesearchnet} to ensure the accurate identification of function names, docstrings, and corresponding code. Tree-Sitter was particularly beneficial for fixing associations between function names and their related data at the file level, enabling precise parsing and confirming the integrity of our dataset.
To avoid duplication, in cases where multiple consecutive updates were made to the same function, we only included the first instance in our dataset. This ensured that each function was represented by a single update, maintaining diversity of the data set. We then applied a 50-50 train-test split, based on a random selection of the curated projects. This split ensures an even distribution of data for training and evaluation. Different components of our dataset presented in ~\autoref{tab:dataset_schema}. Apart from code and docstring this dataset includes commit messages, commit SHA, code diff, and docstring diff.

\begin{table}[h!]
\centering
\begin{tabular}{|l|p{0.6\columnwidth}|}
\hline
\multicolumn{1}{|c|}{\textbf{Field}}               & \multicolumn{1}{|c|}{\textbf{Description}}                                                                                              \\ \hline
\texttt{file}                & Name of the file being analyzed.                                                                                  \\ \hline
\texttt{function}            & Name of the function in the file.                                                                                 \\ \hline
\texttt{version\_data}       &  Each entry contains file-version metadata, in array form, including: \texttt{docstring\_lines}, \texttt{code\_lines}, \texttt{commit\_date\_time}, \texttt{commit\_sha}, \texttt{commit\_message}, \texttt{docstring} and \texttt{code}.               \\ \hline

\texttt{docstring\_lines}    & Version-specific start and end line numbers of the function's docstring.                                                          \\ \hline
\texttt{code\_lines}         & Corresponding line numbers for code.                                                               \\ \hline
\texttt{commit\_date\_time}  & Timestamp of version's commit.                                                             \\ \hline
\texttt{commit\_sha}         & SHA of the commit for a specific version.                                                                                               \\ \hline
\texttt{project}             & Name of the project.                                                                                             \\ \hline
\texttt{owner}               & Owner or organization maintaining the project.                                                                   \\ \hline
\texttt{filename}            & Name of the file.                                                        \\ \hline
\texttt{file\_path}          & Full path to the file.                                                                                           \\ \hline
\texttt{commit\_message}     & Commit message for the specific version.                                                                      \\ \hline
\texttt{docstring}           & The function's docstring for that version.                                                                             \\ \hline
\texttt{code}                & Function source code for that version.                                                                                     \\ \hline
\texttt{diff\_code}          & Unified diff of code changes between versions.                                                                   \\ \hline
\texttt{diff\_docstring}     & Unified diff of docstring changes.                                                              \\ \hline
\end{tabular}
\caption{Schema of the \dsnametitle{} dataset.}
\label{tab:dataset_schema}
\end{table}

\begin{figure}[h!]
    \centering
    \begin{subfigure}[b]{0.45\linewidth}
        \centering
        \includegraphics[width=\linewidth]{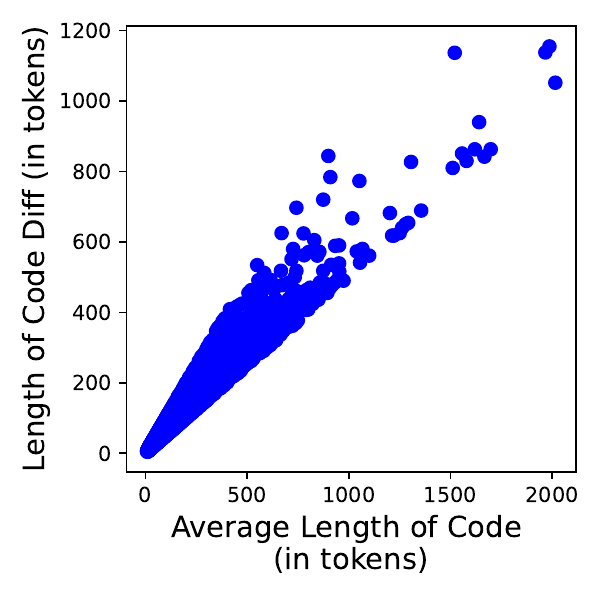}
        \caption{Scatterplot of the length of function code diff  \emph{vs.} the average function length, both in tokens; longer functions tend to have longer diffs.}
        \label{fig:code_scatter}
    \end{subfigure}
    \hfill
    \begin{subfigure}[b]{0.45\linewidth}
        \centering
        \includegraphics[width=\linewidth]{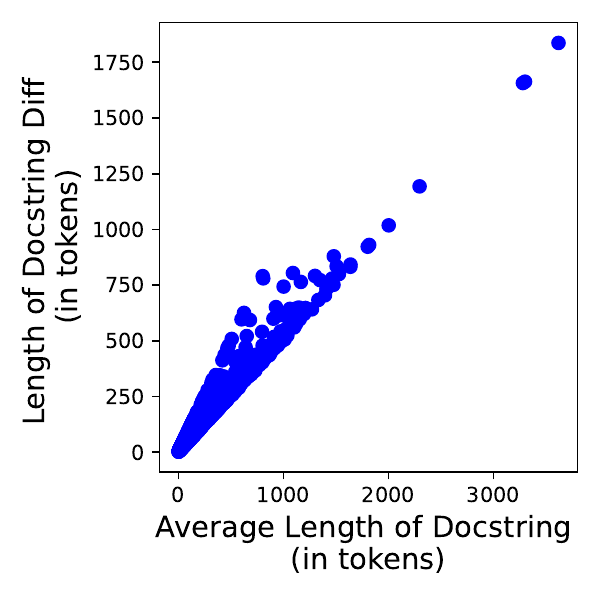}
        \caption{Scatterplot showing the length of docstring diff \emph{vs.} the average docstring length, both in tokens; longer docstrings tend to have longer diffs.}
        \label{fig:docstring_scatter}
    \end{subfigure}
    \caption{Code and docstring diff statistics.}
    \label{fig:combined_scatterplots}
\end{figure}
\begin{figure}[t]
    \centering
    \includegraphics[width=0.75\linewidth]{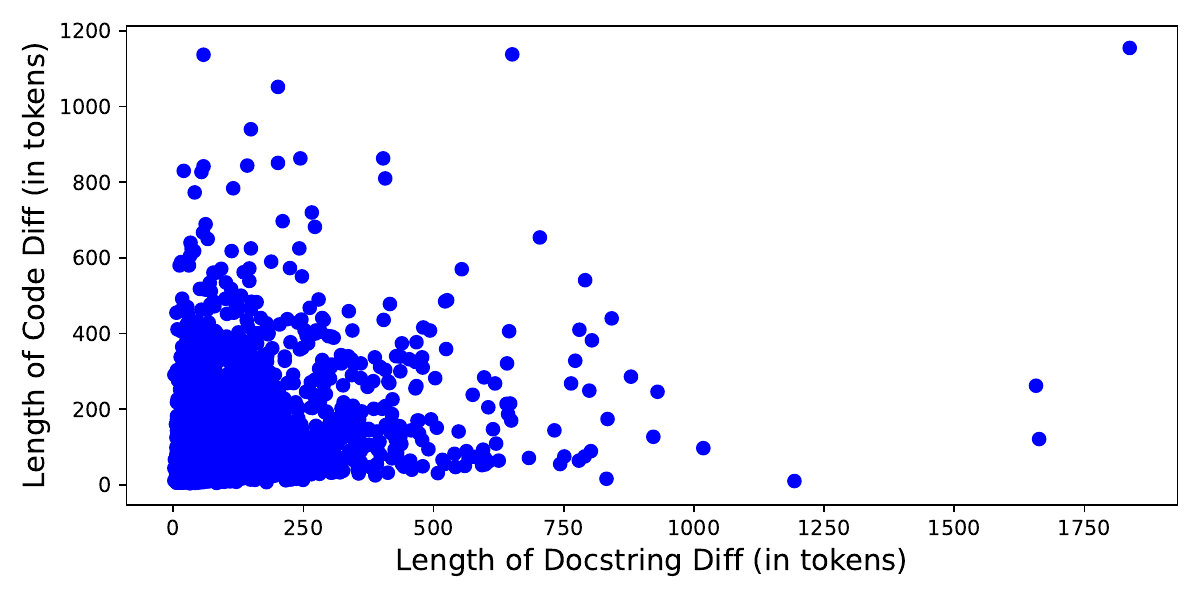}
    \caption{Scatterplot showing length of code diff (in tokens) vs. length of docstring diff (in tokens)}
    \label{fig:diff_code_diff_docstring}
    \vspace{-10pt}
\end{figure}

\noindent\emph{Distribution of Code Changes:}
~\autoref{fig:code_scatter} plots the length of functions, 
\emph{vs} the length of the diffs to those functions (both lengths measured in tokens); the plot suggests that longer functions have longer diffs. 
This also applies to docstring.~\autoref{fig:docstring_scatter} shows the length of the docstring diff vs. the average docstring length in tokens, on a function level.
Longer documentation is more susceptible to significant changes. This indicates the challenges in maintaining consistency in descriptive text, especially as the underlying code evolves.
\autoref{fig:diff_code_diff_docstring} shows the length of the code diff vs. the length of the docstring diff, both in tokens.
There is less relation between the length of the code diff and the length of the docstring diff, indicating difference in the detail of documentation for functions.


\vspace{.1cm}

\begin{figure}
    \centering
    \includegraphics[width=0.75\linewidth]{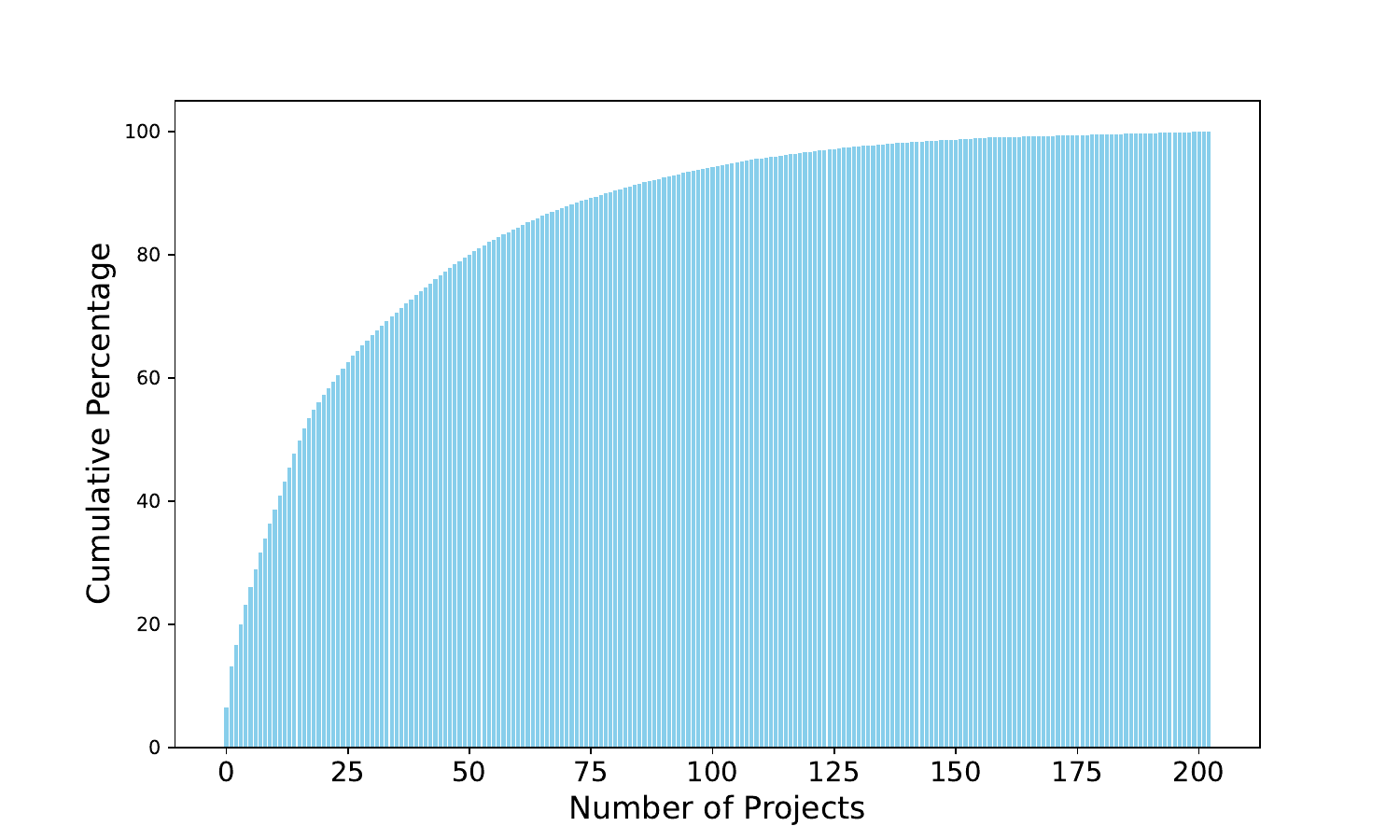}
    \caption{Cumulative percentage distribution of dataset entries by project}
    \label{fig:cumulative_distribution}
    \vspace{-10pt}
\end{figure}

\noindent\emph{Project Distribution:}
Figure \ref{fig:cumulative_distribution} shows a lift chart of the fraction of dataset entries from each project, with the largest projects included leftmost.
As with a lot of software data,  this dataset is skewed, 
with the top 25 projects accounting for around $60\%$ of the total samples.
The remaining $40\%$ of the dataset is spread across the remaining projects.
Our dataset includes a diverse range of projects, providing a broad selection of code and docstring changes across different codebases, ranging from function lengths of 4 lines to 490 lines, and from 1 commit to 301 total commits per project.

\section{Research Questions} 
As a basic illustration of the use of our dataset, 
we investigate the ability of large language models (LLMs) to comprehend and generate aligned updates in code-docstring pairs under the following research questions:

\begin{tcolorbox}[colback=white, colframe=black, sharp corners, title=Research Questions]
    \begin{enumerate}[leftmargin=*]
        \item Can LLM-generated code \& docstrings (for old and new versions) correctly align with the ground (old \& new respectively) truth code \& docstrings?
        \item Can LLMs update code given an updated docstring (or vice versa)?
    \end{enumerate}
\end{tcolorbox}

\noindent\textbf{RQ1} examines whether LLM-generated code
and docstrings can correctly align with the reference old and new version of code or docstrings.
To test this, we test the LLM with code/docstring pairs, 
without indicating their temporal context. For docstring generation, we asked the model to generate docstrings for old and new code, producing two outputs: \texttt{old\_gen\_docstring}, \texttt{new\_gen\_docstring}. We then compare to the two references: \texttt{old\_ref\_docstring}, and \texttt{new\_ref\_docstring},
(gathered from the repo as described above).
For correct alignment, we require that the edit distance~\cite{edit_distance,navarro2001guided,levenshtein1966binary} between \texttt{old\_gen\_docstring} and \texttt{old\_ref\_docstring} should be lower than the distance between
\texttt{old\_gen\_docstring} and \texttt{new\_ref\_docstring}.  
We use edit distance because it is applicable to both code and docstrings. Besides, for many instances, the difference is very small and edit distance can reflect it better.
The process is repeated in reverse: with docstrings as input and code as output.
This task is rather demanding exercise, first of
code summary generation, and then of code generation; for correct alignment, 
these generation tasks require the LLM to respond properly to {\bf \emph{incremental}} changes---in the former case, in code, and in the latter, in the summary. 

\vspace{.1cm}

\noindent\textbf{RQ2} requires LLMs to effectively update code based on an updated docstring (and vice versa). The baseline is that the model is tasked with generating updated code (\texttt{new\_gen\_code}) using the old docstring, old code, and the new docstring as inputs. The generated output is then evaluated against the reference updated code (\texttt{new\_ref\_code}) and the original reference code (\texttt{old\_ref\_code}) using edit distance. A similar process is applied to evaluate the model's ability to update the docstrings. 

We also tried different prompting strategies:

\begin{itemize}
    \item \emph{Incorporating Contextual Information:} We included relevant project-level metadata, such as the project name, owner, file path, and commit message to the input. 
    \item \emph{3-Shot Learning with BM25:} Using BM25~\cite{robertson2009probabilistic}, we retrieve 3 older code snippets from the training set that were most similar to the given old code. These were incorporated as few-shots. 
    \item \emph{Hybrid Strategy:} We use both the above, together, 
    to check if this improves performance. 
\end{itemize}

\vspace{.1cm}

\noindent\emph{The Models:}
We used the Instruct Turbo version of Meta's Llama-3.1, with 405 billion parameters~\cite{dubey2024llama}, and the Instruct v0.1 version of MistralAI's Mixtral, with 22 billion parameters and 8 feedforward blocks per layer~\cite{jiang2024mixtral} for our experiments.
The Mixtral 8x22B~\cite{jiang2024mixtral} model is a sparse mixture-of-experts (SMoE) that dynamically selects and combines two out of eight expert groups per token, allowing it to use 47 billion parameters while only activating 13 billion for each token, making it efficient and highly effective.
In contrast, Meta's Llama 3 series~\cite{dubey2024llama}, employs a dense Transformer architecture. Llama 3 models have improved upon previous generations primarily through improved data quality, greater diversity, and expanded training scale.
Both models are competitive in tasks related to code and natural language generation, making them appropriate for our study.

\section{Results}





\begin{table}[h!]
\centering
\renewcommand{\arraystretch}{1.2}
\setlength{\tabcolsep}{6pt} 
\begin{tabular}{|p{2.5cm}|l|>{\raggedleft\arraybackslash}p{1cm}|>{\raggedleft\arraybackslash}p{1cm}|>{\raggedleft\arraybackslash}p{1cm}|}
\hline
\textbf{Model} & \textbf{Type} & \textbf{Aligned: New} & \textbf{Aligned: Old} & \textbf{Aligned: Both} \\ \hline
\multirow{2}{*}{\parbox[t]{4cm}{Mixtral-8$\times$22B\\ Instruct v0.1}} 
    & Code & 1130 & 1303 & 352 \\ \cline{2-5} 
    & Docstring & 1282 & 949 & 217 \\ \hline
\multirow{2}{*}{\parbox[t]{4cm}{Meta Llama-3.1 405B\\ Instruct Turbo}} 
    & Code & 1191 & 1214 & 407 \\ \cline{2-5} 
    & Docstring & 1304 & 1002 & 331 \\ \hline
\end{tabular}
\caption{Results of RQ1 for Mixtral-8$\times$22B and Meta Llama-3.1 405B (both zero-shot) on 2273 test samples. For example, ``Aligned: New'' refers to the number of outputs better aligned with the new reference. See text for additional explanation.}
\label{table:results-rq1}
\end{table}


\begin{table}[h!]
\centering
\renewcommand{\arraystretch}{1.1}
\setlength{\tabcolsep}{4pt}

\begin{tabular}{|p{3cm}|p{2.5cm}|p{1.25cm}|r|}
\hline
\textbf{Method} & \textbf{Model} & \textbf{Type} & \textbf{Correct} \\ \hline

\multirow{4}{*}{0-shot} 
& \multirow{2}{*}{\parbox[t]{4cm}{Mixtral-8$\times$22B\\ Instruct v0.1}} & Code & 751 \\ \cline{3-4} 
&  & Docstring & 1255 \\ \cline{2-4} 
& \multirow{2}{*}{\parbox[t]{4cm}{Meta Llama-3.1 405B\\ Instruct Turbo}} & Code & 1081 \\ \cline{3-4} 
&  & Docstring & 1001 \\ \hline

\multirow{4}{*}{\parbox[t]{4cm}{0-shot w/\\ Contextual Information}} 
& \multirow{2}{*}{\parbox[t]{4cm}{Mixtral-8$\times$22B\\ Instruct v0.1}} & Code & 909 \\ \cline{3-4} 
&  & Docstring & 1304 \\ \cline{2-4} 
& \multirow{2}{*}{\parbox[t]{4cm}{Meta Llama-3.1 405B\\ Instruct Turbo}} & Code & 1096 \\ \cline{3-4} 
&  & Docstring & 1087 \\ \hline

\multirow{4}{*}{3-shot w/ BM25 Retrieval} 
& \multirow{2}{*}{\parbox[t]{4cm}{Mixtral 8$\times$22B\\ Instruct v0.1}} & Code & 644 \\ \cline{3-4} 
&  & Docstring & 1233 \\ \cline{2-4} 
& \multirow{2}{*}{\parbox[t]{4cm}{Meta Llama-3.1 405B\\ Instruct Turbo}} & Code & 1127 \\ \cline{3-4} 
&  & Docstring & 921 \\ \hline

\multirow{4}{*}{\parbox[t]{4cm}{3-shot w/ BM25 Retrieval \\ \& Contextual Information}} 
& \multirow{2}{*}{\parbox[t]{4cm}{Mixtral-8$\times$22B\\ Instruct v0.1}} & Code & 785 \\ \cline{3-4} 
&  & Docstring & 1311 \\ \cline{2-4} 
& \multirow{2}{*}{\parbox[t]{4cm}{Meta Llama-3.1 405B\\ Instruct Turbo}} & Code & 879 \\ \cline{3-4} 
&  & Docstring & 969 \\ \hline

\end{tabular}
\caption{Results of RQ2 for Mixtral-8$\times$22B and Meta Llama-3.1 405B on 2273 test samples}
\label{table:results-rq2}
\end{table}

The results presented in Tables \ref{table:results-rq1} and \ref{table:results-rq2} highlight the performance of Mixtral-8$\times$22B and Meta Llama-3.1 405B.
The tables show the number of correct alignments for each model and method, including 0-shot, 0-shot with contextual information, 3-shot with BM25 retrieval, and 3-shot with both BM25 retrieval and contextual information.

For RQ1, we define a correct alignment as: the generated code/docstring from the old docstring/code has a lower edit distance to the old code/docstring than the new code/docstring \emph{and} the generated code/docstring from the new docstring/code had a lower edit distance to the new code/docstring than the old code/docstring, implying perfect temporal alignment. For RQ2, we define a correct alignment as: the generated code/docstring from the new docstring/code pair had a lower edit distance to the new (reference) code/docstring than the old code/docstring, as we are using the old docstring-code pair as contextual information to help update the code/docstring. Our rationale is that lower edit distance suggests that
the model's generation actually saves developers some editing work; future work could use other metrics. 


Results in~\autoref{table:results-rq1} shows the correctly aligned sample counts (rightmost column); other columns show counts of right alignment for just the old and new code or docstrings. 
The test set includes $2273$ samples.
For RQ1 (and RQ2) the models struggle, indicating that this is a
challenging task. 
The best performer ($407$ correct identifications) is  Meta Llama-3.1 405B in RQ1 and $1311$ correct identifications for Mixtral-8$\times$22B in RQ2 (around $18\%$ and $58\%$ of the total samples, respectively).
Just considering the alignment of code and docstring to old references and new references separately for RQ1:  models struggle to achieve high accuracy, with the best performance being $1303$ correct identifications for Mixtral-8$\times$22B, which is around $57\%$ of the total samples (slightly better than a coin flip).
The results suggest more room for improvement in understanding and generating aligned updates in code-docstring pairs.

For RQ2 (\autoref{table:results-rq2}), the models are better at docstring updates, compared to code updates.
Adding contextual information improves the alignment of the models towards the new references; but 
using a 3-shot prompting setup with BM25 retrieval does not help,
except for Meta Llama-3.1 405B in the case of code updates.
Adding contextual information to the 3-shot learning setup improves performance  somewhat; however, while improving over baseline,  it is still worse than using contextual information alone, except in the case of Mixtral-8$\times$22B for docstring updates which slightly outperformed the 0-shot with contextual information setup with $1311$ correct identifications compared to $1304$.

Another interesting result is that Mixtral-8$\times$22B largely performs better with the 3-shot learning setup compared to Meta Llama-3.1 405B, which suggests that the model is better at learning from examples.

\section{Limitations}
A key limitation of the dataset is its inability to track changes to a function if the function is moved to a file with a different name.
This limitation arises because the tracking mechanism relies on the file name remaining consistent.
Moreover, if the file name changes or the function is relocated to another file, the dataset cannot accurately trace its modifications or evolution over time.
This constraint may impact the dataset's utility in scenarios where file restructuring or renaming is common.
Additionally, the dataset is limited to tracking changes within the main branch only.
This limitation was implemented to ensure consistency and stability in the tracked code and docstring.
The main branch is typically considered the primary branch for the most stable and production-ready version of the code.
By focusing on this branch, the dataset avoids potential complications arising from the variability and experimental nature of feature branches or pull requests, which may not always reflect high-quality code and docstring often seen in the final, stable codebase.

\section{Conclusion}
Our source code is publicly available on Zenodo and GitHub~\cite{Pai_CoDocBench_A_Dataset_2024}. The link to the DOI-identifier is: \href{https://doi.org/10.5281/zenodo.14251622}{https://doi.org/10.5281/zenodo.14251622}

\section*{Acknowledgments}
This work was partially supported by the National Science Foundation under CISE SHF MEDIUM 2107592.






%
\bibliographystyle{IEEEtran}
\bibliography{references}

\end{document}